\input{aipcheck}

\documentclass[
    ,final            
  ]
  {aipproc}

\layoutstyle{6x9}

\begin{document}

\title{A possible signature of primordial stellar populations in $z=3$
Lyman $\alpha$ emitters}

\classification{<Replace this text with PACS numbers; choose from this list:
                \texttt{http://www.aip..org/pacs/index.html}>}

\author{Akio K. Inoue}{
  address={College of General Education, Osaka Sangyo University,
  3-1-1, Nakagaito, Daito, Osaka 574-8530, Japan}
}

%

\begin{abstract}
Observations with Subaru telescope have detected surprisingly strong
 Lyman continuum (LyC; $\sim900$ \AA\ in the rest-frame) from some Lyman
 $\alpha$ emitters (LAEs) at $z=3.1$. We have examined the stellar
 population which simultaneously accounts for the strength of the LyC
 and the spectral slope of non-ionizing ultraviolet of the LAEs. As a
 result, we have found that stellar populations with metallicity
 $Z\geq1/50~Z_\odot$ can explain the observed LyC strength only with a
 very top-heavy initial mass function (IMF; 
 $\langle m \rangle \sim 50 M_\odot$). However, the critical metallicity
 for such an IMF is expected to be much lower. A very young ($\sim1$
 Myr) and massive ($\sim100$ $M_\odot$) extremely metal-poor
 ($Z\leq5\times10^{-4}Z_\odot$) or metal-free (so-called Population III)
 stellar population can also reproduce the observed LyC strength if the
 mass fraction of such `primordial' stellar population is
 $\sim1$\% in total stellar mass of the LAEs.
\end{abstract}

\maketitle


\section{Introduction}

The first generation of stars in the Universe is the stellar
population without any metal elements. Although the formation of such
metal-free stars, or so-called Population III (Pop III) stars may last
until $z\sim2$--3 if the metal enrichment in the intergalactic medium
(IGM) is inefficient, there is no firm observational evidence of the
stellar population.

The stellar population with metallicity $Z<10^{-5}$, which is $<1/2000$
$Z_\odot$, is classified as extremely metal-poor (EMP) stars. Low-mass
EMP stars found in the halo of the Galaxy are survivors of the early
stage of the formation of the Galaxy. On the other hand, it is expected
that their high-mass counterpart existed in the early days and died out
until the current epoch. Yet, there is no direct observational evidence
of such massive EMP stars at high-$z$.

This paper presents a possible signature of such `primordial' stellar
populations in high-$z$ galaxies. There is a new population of galaxies
at $z\sim3$ detected in their rest-frame $\sim900$ \AA, Lyman continuum
(LyC). Surprisingly strong LyC relative to non-ionizing ultra-violet (UV)
was reported by \cite{Iwata2009}. This may indicate the presence of the
primordial stellar populations as shown in \cite{Inoue2010,Inoue2011a}
and below.

\section{Rest UV two-colour diagram}

\begin{figure}
  \includegraphics[width=0.9\textwidth]{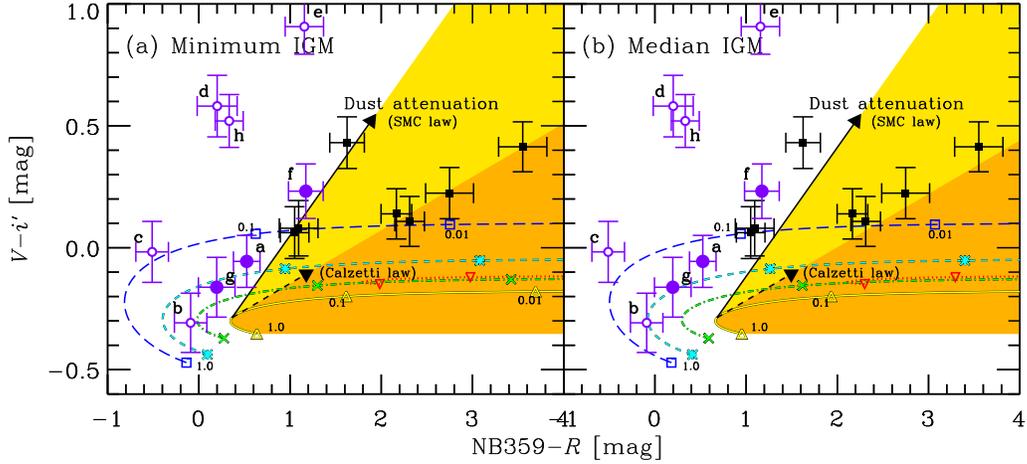}
  \caption{Rest UV two-colour diagram for $z\simeq3.1$ LyC emitting
 galaxies. The vertical axis, $V-i'$, indicates non-ionizing UV spectral
 slope, and the horizontal axis, NB359$-R$, indicates LyC-to-UV flux
 density ratio. The filled circles with errorbars are the 3 LAEs without
 Ly$\alpha$ line offset \cite{Inoue2011a} and the open circles with
 errorbars are the 5 LAEs with the line offset \cite{Inoue2011a}. The
 filled squares with errorbars are LBGs reported in \cite{Iwata2009}. 
 The colours expected from the SED model with escaping stellar and
 nebular emissions developed by \cite{Inoue2010} are shown by the
 dotted curve with inverse triangles (model Ac in Table 1), the solid
 curve with triangles (model A), the dot-dashed curve with x-marks
 (model B), the short-dashed curve with asterisks (model C), and the
 long-dashed curve with squares (model D). These curves are a function
 of $f_{\rm esc}$ and the colours for $f_{\rm esc}=1.0$, 0.1, and 0.01
 are indicated by the symbols. We have assumed minimum and median IGM
 attenuations in panels (a) and (b), respectively, based on
 \cite{Inoue2008,Inoue2011a}. The two diagonal arrows show dust
 attenuations with $E(B-V)=0.1$ for the SMC law (solid arrow) and the
 Calzetti law (dashed arrow). The shaded regions indicate the regions
 explained by the stellar population model A with a combination of IGM
 and dust attenuations.}
\end{figure}

\cite{Iwata2009} and \cite{Inoue2011a} have found extremely strong
LyC from some Lyman $\alpha$ emitters (LAEs) at $z=3.1$. Their redshifts
were confirmed with spectroscopy. However, some LAEs show a spatial
offset between Ly$\alpha$ emission line and LyC. These objects may have
a faint foreground galaxy accounting for the `pseud-LyC'. However, it is
difficult to attribute all the objects to foreground contamination in a
statistical sense because of the smallness of the offset ($<1''$)
\citep{Inoue2011a}. Figure~1 shows the extreme strength of the LyC of
the LAEs found by \cite{Iwata2009,Inoue2011a}. The filled circles are
the most reliable objects without any offset, while the open circles are
the objects with the small offset. The filled squares are the sample of
Lyman break galaxies of \cite{Iwata2009}.

\subsection{Single stellar population model}

\begin{table}
 \begin{tabular}{lccccc}
   \hline
   Model & $Z/Z_\odot$ & IMF & SF history & age & SED reference\\
   \hline
   Ac & 1/5 & Salpeter & Constant & 100 Myr & {\sc starburst}99 (v.5.1)\\
   A & 1/50 & Salpeter & Instantaneous & 1 Myr & {\sc starburst}99 (v.5.1)\\
   B & 1/50 & Top-heavy & Instantaneous & 1 Myr & {\sc starburst}99 (v.5.1)\\
   C & 1/2000 & Massive & Instantaneous & 1 Myr & Schaerer (2003)\\
   D & 0 & Massive & Instantaneous & 1 Myr & Schaerer (2003)\\
   \hline
 \end{tabular}
 \caption{Stellar population models.}
 \label{tab:a}
\end{table}

Let us compare the observed colours with a spectral model with a
single stellar population. In particular, we allow an escape of nebular
LyC (bound-free recombination continuum) as well as an escape of stellar
LyC as described in \cite{Inoue2010}. This escaping nebular LyC boosts
the flux just below the Lyman limit. We call it `Lyman limit bump'
\citep{Inoue2010}. Actually, the NB359 filter used in observations of
\cite{Iwata2009} exactly traces the Lyman limit bump at $z=3.1$. Then,
the NB359$-R$ colour becomes bluest not when the LyC escape fraction 
$f_{\rm esc}=0$ but when $f_{\rm esc}\approx0.5$. This results in a
round shape of the model curves in Figure 1 which are a function of 
$f_{\rm esc}$. Among 5 types of the stellar populations in Table 1, we
find that the stellar populations with an ordinary sub-solar metallicity
and with a Salpeter IMF (i.e. models A and Ac) do not fit the observed
LAEs. If we insist on an ordinary metallicity, the IMF should be massive 
(models B). However, such a massive IMF is expected only at extremely
low metallicity like $Z<10^{-(3-6)}$ $Z_\odot$ \cite{Schneider2006}.
If we wish to explain the LAEs bluest in NB359$-R$, some but not all of
which may be foreground contamination, a stellar population with massive
EMP or metal-free (models C or D) is required.

\subsection{Two stellar population model}

\begin{figure}
  \includegraphics[width=0.9\textwidth]{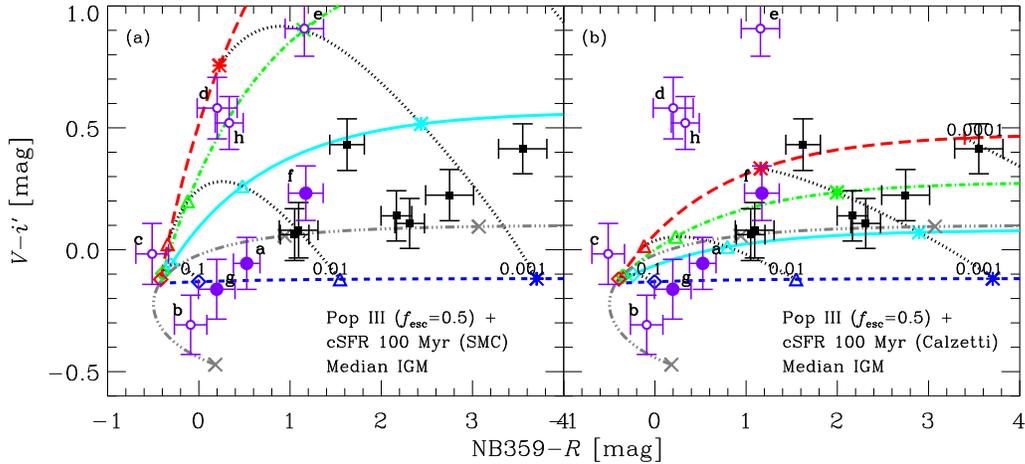}
  \caption{Same as Fig.~1 but comparisons with two stellar population
 models: the underlying normal stellar population (model Ac in 
 Table 1) and the additional Pop III stars (model D). The SMC dust
 extinction law is assumed for the panel (a) but the Calzetti
 attenuation law is assumed for (b). The median IGM attenuation is
 assumed for both panels. The short-dashed curves are the sequences of
 the colour as a function of the mass fraction of Pop III stars for
 dust-free underlying stellar population. The solid, dot-dashed, and
 long-dashed curves are the same sequences but for dusty underlying
 population with $E(B-V)=0.1$, 0.2, and 0.3, respectively. Note that the
 Pop III stars are assumed to be always dust-free. The positions for the
 mass fraction of 0.1, 0.01, 0.001, and 0.0001 are indicated by dotted
 curves with diamonds, triangles, asterisks, and plus-mark,
 respectively. For Pop III stars, the contribution of escaping nebular
 LyC is taken into account, assuming $f_{\rm esc}=0.5$. The colour
 sequence with other $f_{\rm esc}$ is shown by the triple-dot-dashed
 curve as in Fig.~1. For the underlying population, $f_{\rm esc}=0.01$
 is assumed.}
\end{figure}

The previous subsection shows that the LAEs detected in LyC by
\cite{Iwata2009} are likely to contain a primordial stellar population
such as massive Pop III or EMP stars. But how much amount of these
exotic stars are required in them? To discuss these questions, let
us consider a system in which a primordial stellar population and a
normal stellar population with sub-solar metallicity and Salpeter IMF
coexist. We assume that the normal population makes stars with a
constant rate; the model Ac in Table~1. 
For the primordial population, we adopt the model D (Pop III). The mass
fraction of the primordial population in the total stellar mass is a
parameter for the mixture. The normal population may exist in dusty
environment, but the primordial population are likely to be
dust-free. A result is shown in Figure 2.

The LAEs without offset (objects {\bf a}, {\bf f} and {\bf g}; filled
circles in Fig.~2) require a primordial mass fraction of 0.1--10\%. 
The bluest LAEs (objects {\bf b} and {\bf c}; open circles) require a
primordial mass fraction of more than 10\%, but we could not reject the
possibility that their NB359 flux was foreground contamination
individually \cite{Inoue2011a}. The LAEs which are red in $V-i'$
(objects {\bf d}, {\bf e} and {\bf h}; open circles) can be explained by
a model with the SMC law and a small amount of attenuation as
$E(B-V)\simeq0.2$--0.3 and with a primordial mass fraction of a few
0.1\%. Note that the possibility of foreground contamination, especially
for {\bf d} and {\bf e}, is the largest because of the largest line
offset \cite{Inoue2011a}.

\section{Summary and future prospects}

We have suggested that the extreme strength of LyC observed in some LAEs
at $z=3.1$ indicates the presence of massive EMP or Pop III stars in
them. Such a galaxy population could be a cosmic `ionizer' for the
reionization at $z>6$. Indeed, there seems an evolution of LyC
emissivity of galaxies toward high-$z$ \citep{Inoue2006}. This may be
caused by an increase of this type of the galaxy population at
high-$z$. In future, we should confirm the presence of the primordial
stellar population in the galaxies. For example, we can constrain the
metallicity to be $Z<10^{-3}$ $Z_\odot$ if the ratio of [OIII]
$\lambda$5007 to H$\beta$ is less than 0.1 \citep{Inoue2011b}. Deep
near-infrared spectroscopy for the LAEs is strongly encouraged.


\begin{theacknowledgments}
The author thanks K. Kousai, I. Iwata, Y. Matsuda, E. Nakamura, 
M. Horie, T. Hayashino, C. Tapken, M. Akiyama, S. Noll, 
T. Yamada, D. Burgarella and Y. Nakamura for collaboration in this
research.
\end{theacknowledgments}



\bibliographystyle{aipproc}   




\end{document}